\documentclass[journal]{IEEEtran}
\ifCLASSINFOpdf
  \usepackage[pdftex]{graphicx}
\else
  \usepackage[dvips]{graphicx}
\fi
\usepackage{amsmath}
\usepackage{tabularx}
\usepackage{booktabs}
\usepackage{lscape}
\usepackage{longtable}
\usepackage{lipsum}
\usepackage[dvipsnames]{xcolor}
\definecolor{myblue}{rgb}{0.36, 0.54, 0.66}
\definecolor{mygreen}{rgb}{0.53, 0.66, 0.42}
\definecolor{mypurple}{rgb}{0.6, 0.4, 0.8}
\definecolor{mypink}{RGB}{219, 112, 147}
\usepackage{multirow}
\usepackage{cite}

\usepackage{algorithmic}

\usepackage{array}
\newcolumntype{P}[1]{>{\centering\arraybackslash}p{#1}}
\begin{document}
%
\title{Electric Vehicle Charging Load Modeling: A Survey, Trends, Challenges and Opportunities}
%
%
%

\author{
Xiachong~Lin,
Arian Prabowo,
Imran Razzak,
Hao Xue,
Matthew Amos,
Sam Behrens, \\
Flora D. Salim
\thanks{X. Lin, A. Prabowo, I. Razzak, H. Xue, F. Salim are with University of New South Wales, Sydney}
\thanks{M. Amos, S. Behrens are with CSIRO}
}
\maketitle

\begin{abstract}
The evolution of electric vehicles (EVs) is reshaping the automotive industry, advocating for more sustainable transportation practices. Accurately predicting EV charging behavior is essential for effective infrastructure planning and optimization. However, the charging load of EVs is significantly influenced by uncertainties and randomness, posing challenges for accurate estimation. Furthermore, existing literature reviews lack a systematic analysis of modeling approaches focused on information fusion. This paper comprehensively reviews EV charging load models from the past five years. We categorize state-of-the-art modeling methods into statistical, simulated, and data-driven approaches, examining the advantages and drawbacks of each. Additionally, we analyze the three bottom-up level operations of information fusion in existing models. We conclude by discussing the challenges and opportunities in the field, offering guidance for future research endeavors to advance our understanding and explore practical research directions.
\end{abstract}


%
\IEEEpeerreviewmaketitle

\section{Introduction}
\begin{figure}[!htb]
    \centering
    \includegraphics[width=9cm]{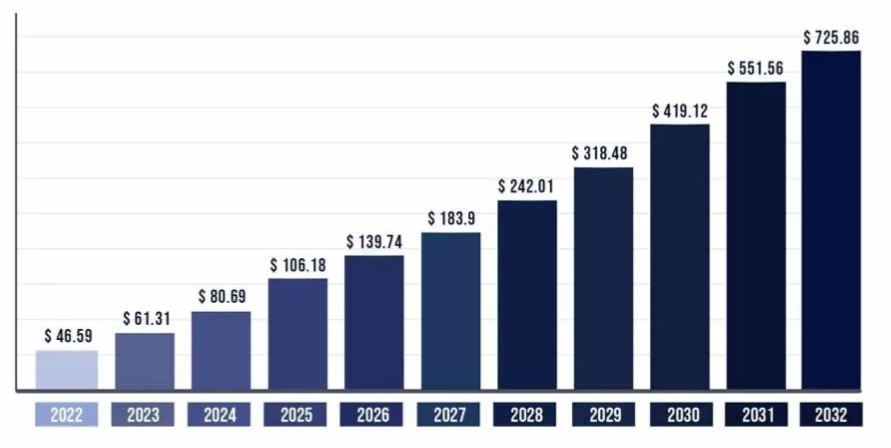}
    \caption{EVs charging station market size from 2023 to 2032$^1$}
    \label{fig:market_size}
\end{figure}

\IEEEPARstart{T}{he} global automotive landscape is undergoing a profound transformation as electric vehicles (EVs) emerge as a pivotal force in the drive towards a more sustainable and environmentally conscious transportation system. Nowadays, EVs are becoming increasingly popular as they offer a cost-effective alternative to traditional internal combustion engine vehicles with better driving experience. Its integration into the modern energy landscape is not only reshaping the automotive industry but also challenging the conventional boundaries of the power grid, which is significant for pushing sustainable transportation and developing smart cities. With significant attention from governments, automakers, and consumers, the EV industry is projected to hit around USD 725.86 billion by 2032, growing at a CAGR of 31.6\% (Figure \ref{fig:market_size} shows the market size from 2023 to 2032\footnote{https://www.novaoneadvisor.com/report/sample/6686}).

 As EV adoption continues to rise globally, operations towards EV charging and infrastructure management, including load monitoring, infrastructure planning, and strategy optimization, become increasingly critical. Accurate estimation of the EV charging behavior serves as the foundation for the above-mentioned engineering purposes and has received rising attention from academic and industrial communities. However, estimating EV charging sessions is challenging due to the randomness and uncertainties influenced by natural, eco-social, and human psychological factors. These factors involve information from engineering knowledge, physics law, and relevant datasets. How to combine information from knowledge, data, and outputs from various sources or processes, such as sensors, algorithms, models, or human experts, to generate a comprehensive and accurate representation of the charging sessions is a significant bottleneck waiting for further open discussion. Existing reviews in the context of EV charging concentrate on popular techniques reviewed, as well as impact analysis based on grid regulation and market operations. Amara et al. \cite{amara2021review, amara2021review_poster} provided statistics for available open datasets with historical charging sessions (HCS) and reviews about EV charging load models. Xiang et al. \cite{xiang2019electric} concentrated on simulation-based methods, analyzing the factors influencing the EV user behavior and charging load in the power grid. Machine learning approaches utilized in EV charging modeling were reviewed by Shahriar et al. \cite{shahriar2020machine}. Another survey about EV charging management based on reinforcement learning was provided by Abdullah et al. in reference \cite{abdullah2021reinforcement}. In addition, the positive and negative impacts of electric vehicle integration on the power grid and the potential of the grid revolution were analyzed in \cite{tavakoli2020impacts, mohammad2020integration, nour2020review, jia2023integrating}. 
This paper presents a systematic literature review of EV charging load modeling by studying the up-to-date EV datasets, resources, and methods. The main contributions of this paper include:

\begin{itemize}
    \item This survey offers a structured review of existing works, examining their design scope, utilized datasets, and employed modeling approaches. It categorizes the modeling methods and provides a detailed analysis of their strengths and limitations. 
    \item The paper provides a comprehensive overview of recent literature on EV charging load modeling, offering valuable insights into the evolving methodologies and applications in the field. It highlights the key features and innovations found in the most up-to-date studies.
    \item Additionally, this review identifies the challenges, trends, and opportunities in the field of EV charging load modeling. It highlights gaps in current research, helping researchers recognize areas for further exploration and improvement. The survey also sheds light on emerging trends, offering inspiration for future research and innovative approaches to address the identified challenges.
\end{itemize}

\section{Problem Formulation and Research Framework}
\subsection{Problem Formulation}
EV charging load modeling serves various purposes, aiding in EVs' efficient and sustainable integration into the energy landscape. Supportive engineering application encompassing:
\begin{itemize}
    \item \emph{Load Monitoring and Management:} EV charging load models help grid operators monitor electricity consumption patterns and predict the impact of EV charging on the power grid, ensuring sustainable integration.
    \item \emph{Infrastructure Planning:} City planners and policymakers use EV charging load models to strategically plan the placement and capacity of charging stations or propose a charging strategy installed in the digital infrastructure system, facilitating the development of an effective charging network.
    \item \emph{Grid Operation and Regulation:} EV charging load models assist grid operators in balancing supply and demand, maintaining grid stability, and informing policy decisions to regulate the integration of EVs into the grid.
\end{itemize}
The quantitative tasks involved in EV charging load modeling encompass several key aspects aimed at enhancing the understanding and management of EV charging dynamics:
\begin{itemize}
    \item \emph{Demand prediction:} Based on historical data, forecast future EV charging trends, enabling better resource planning and grid management.
    \item \emph{Data augmentation:} Enhance existing datasets by filling in missing information or generating synthetic data, improving the robustness of predictive models.
    \item \emph{Strategy optimization:} Finding the best solution from a set of feasible solutions, typically to maximize or minimize an objective function while satisfying certain constraints. Typical objectives in EV charging load include minimizing the grid loss, voltage variation, and operational costs and maximizing the stakeholder benefits. 
\end{itemize}
Based on the literature review, the scope of EV charging load modeling can be categorized into two main horizons:
\begin{itemize}
    \item \emph{Short-term: } Pertaining to tasks with durations less than one year.
    \item \emph{Long-term: } Encompassing tasks with durations exceeding one year.
\end{itemize}
The spatial dimensions of EV charging load can be categorized into the following levels:
\begin{itemize}
\item \emph{Building-Level: } This pertains to the charging load within an individual or small-scale residential/commercial structure, such as a particular apartment, business establishment, or campus.
\item \emph{Regional-Level: } This encompasses the aggregated electricity load across a broader area, such as a city or province.
\end{itemize}
\subsection{Research Framework and Paper Organization}
To declare the credibility of the research and enhance its value to the community, we illustrated the methodology for conducting the research in this section. Figure \ref{fig:mindmap} provides a mind map that illustrates the backbone of this research. The backbone of this survey is surrounding four research questions: 1) What information do we have? 2) What mechanism do we have for modeling? 3) How do we achieve information fusion based on the available tools? 4) What are the challenges and future trends? We arrange an independent section to respond to each research question, as Section \ref{sec:infowehave}-\ref{sec:discussion}. To comprehensively access states-of-arts, we set up three groups of searching keywords corresponding to Section \ref{sec:infowehave}-\ref{sec:howtointegrate} for paper retrieval. We use static keywords, including 'EV charging', 'EV charging load', 'EV charging load modeling/forecasting', distinctly incorporating three sets of dynamic keywords, which are 
\begin{itemize}
    \item 'open data', 'dataset'
    \item 'spatio-temporal', model name
    \item 'multisource', 'multiprocess', 'multimodel', 'information fusion.'
\end{itemize}
to filter literature on Google Scholar's platform. The time scope of this research is set from Jan 2020 to Jan 2024 since we aim to understand what state-of-the-art is concerned with. According to the analysis of retrieved literature, we conducted a discussion in Section \ref{sec:discussion} and provided a conclusion in \ref{sec:conclusion}.

\section{Information Sources in EV Charging Activities}
\label{sec:infowehave}

EV charging emerges as the outcome of various interconnected and complex ecosystems. It involves information from multiple sources, including the electric grid, aggregators, and EV drivers, all of which contribute to shaping charging behavior. In the modeling process, this information can play two roles: incorporate explicitly, providing numerical data that reflects distribution patterns, or implicitly, influencing the modeling process through feature selection and decision-making.

The information involved in EV charging modeling can be broadly categorized into domain knowledge and digital data. Domain knowledge refers to the expertise and insights derived from industry practices, expert opinions, and established rules. This includes fundamental physical principles, such as battery characteristics, and practical experience related to policy frameworks and regulations (e.g., incentives, regulatory standards, and safety guidelines), as well as market dynamics (e.g., charging costs and EV market penetration). A detailed breakdown of these aspects is provided in Table \ref{tab:domain_knowledge_des}. On the other hand, digital data encompasses large datasets collected from sensors, meters, and user activities, offering valuable insights into charging patterns, energy consumption, vehicle performance, and environmental conditions. Commonly used digital resources and their applications are listed in Table \ref{tab:data}. These digital datasets provide a comprehensive source of information, enabling researchers, policymakers, and industry stakeholders to access data tailored to their specific needs.

In EV charging modeling, domain knowledge and digital data serve complementary functions essential for building accurate, reliable, and adaptable models. Domain knowledge provides the conceptual framework needed to interpret and utilize data effectively, while digital data ensures access to diverse sources, facilitating comprehensive analysis and informed decision-making. The integration of these elements forms a foundation for robust models capable of adapting to technological advancements and evolving market conditions.

\begin{table}[!htb]
    \centering
    \vspace{0.2cm}
    \begin{tabularx}{0.48\textwidth}{lX}
    \toprule
         \textbf{Type} & \textbf{Description} \\
         \midrule
         Battery Characteristics & Knowledge of battery capacities, charging rates, degradation patterns, and thermal dynamics. \\
         \midrule
         Power Grid Dynamics &  Knowledge of grid stability, load balancing, peak demand management, and the impact of EV charging on the grid. \\
         \midrule
         Incentives and Regulations & Government policies promoting EV adoption, incentives for using renewable energy, and regulations governing charging infrastructure. \\
         \midrule
         Safety Standards & Standards and protocols for EV charging equipment and installations. \\
         \midrule
         Cost Models & Understanding the economics of EV charging, including costs associated with electricity, infrastructure, and maintenance. \\
         \midrule
         EV Penetration & Trends in EV adoption rates and market dynamics. \\ 
    \bottomrule
    \end{tabularx}
    \vspace{0.2cm}
    \caption{Detailed Description of Common Used Domain Knowledge in EV Charging Modeling.}
    \label{tab:domain_knowledge_des}
\end{table}

\begin{table*}[!htb]
    \centering
    \vspace{0.2cm}
    \begin{tabularx}{\textwidth}{lXX}
    \toprule
         \textbf{Data Type} & \textbf{Description} & \textbf{Example} \\
         \toprule
         Travel Surveys & Describe the household travel behavior, including origin-destination pairs and hourly trip data for optimizing EV charger locations. & Household Travel Surveys (NHTS) 2009~\cite{NHTS2009}, 2017~\cite{NHTS2017}, 2022~\cite{NHTS2022}. \\ 
         \midrule
         Charging Check-In Data & Detailed information about each charging session in wireless meters monitored charging station, including plug-in/plug-out timestamps, energy uptake, charging cost, charging duration, vehicle identity, etc. & US Department Of Energy EV charging dataset~\cite{DVN/NFPQLW_2020}, Level3 EV charging dataset~\cite{level3_ev_charging_dataset}.\\ 
         \midrule
         Energy Consumption Data & Provides an overview of the regional charging equipment' electricity consumption and energy performance. & Ausgrid electricity usage dataset~\cite{Ausgrid}, NZ Green Grid household electricity demand study \cite{NZElectricityDemand}, EV charging energy usages in Palo Alto~\cite{paloalto}, Dundee~\cite{dundee}, Perth~\cite{perth}, and Boulder Colorado~\cite{boulder}.\\
         \midrule
         Distribution Grid Data & Describe the impact of charging events (e.g., voltage fluctuations) on the grid. & Grid load profiles (IEEE-13, IEEE-30, IEEE-33), power quality data, and renewable energy integration.  \\
         \midrule
         Cost \& Pricing Data & Data indicates the electricity price and charging costs, such as time-of-use tariffs, dynamic pricing, and EVSE charging fee. & California ISO Locational Marginal Price~\cite{CAprice}.\\
         \midrule
         EV Registration Data & Provide the information on EV scale/penetration, the adoption rate of charging modes, statistics about vehicle manufacturer market share, etc. & New Zealand Fleet Statistics~\cite{NZfleet}. \\
         \midrule
         Geographical Data & Provide environmental information about the region map and the area functionality of the charging station. & Beijing Third Ring, Chengdu Third Ring, Sioux Falls Road Network. \\ 
         \midrule
         Meteorological Data & Provides regional weather conditions that affect driving and charging behavior, such as temperature, precipitation, wind speed, humanity, and extreme weather events. & Korea Meteorological Administration (KMA) data~\cite{kma_forecast}, NASA Prediction of Worldwide Energy Resources~\cite{nasa_power}. \\
         \midrule
         Traffic Trajectory & Real-time or historical data on traffic patterns and vehicle movements. & \\
         \midrule
         Comprehensive & Aggregated dataset compassing multi-source data from EV drivers, grid, and EVSE, providing comprehensive insights about EV charging activities. & Data-driven human-centric EV charging dataset~\cite{nie2023data} \\
         \bottomrule
    \end{tabularx}
    \vspace{0.2cm}
    \caption{Detailed description of commonly used digital data in EV charging modeling.}
    \label{tab:data}
\end{table*}

\section{EV Charging Models}
\label{sec:toolswehave}
Differentiated by the fundamental logic of model establishment as shown in Table \ref{tab:fundametaldiff}, the existing approaches modeling the EV charging sessions can be categorized into three types: 1) statistical, 2) simulated, and 3) data-driven. Each category employs distinct methodologies and approaches to capture and predict charging behavior. In this section, we aim to delve deeper into the characteristics of these modeling types, systematically examining their strengths and weaknesses. Through this analysis, we seek to provide insights into the suitability of each approach for specific scenarios, shedding light on the diverse applications within the realm of EV charging infrastructure. By understanding the nuances of statistical, simulated, and data-driven models, stakeholders can make informed decisions on selecting the most appropriate modeling tool based on their charging system's specific requirements and objectives.
\begin{table}[!htb]
    \centering
    \vspace{0.2cm}
    \begin{tabularx}{0.5\textwidth}{lX}
    \toprule
         \textbf{Method} & \textbf{Way to Model Charging Sessions} \\
         \midrule
         Statistical method & Generate joint probabilistic distribution for charging session/demand \\
         \midrule
         Simulated method & Repeatedly sample from the estimated parameter probabilistic distributions and get aggregated results \\
         \midrule
         Data-driven method & Establish predictor and learn its optimal probability parameter $p(\theta|D)$ inherent in the observation data \\
         \bottomrule
    \end{tabularx}
    \vspace{0.2cm}
    \caption{The fundamental differences between statistical, simulated, and data-driven methods in EV charging load modeling. }
    \label{tab:fundametaldiff}
\end{table}
\begin{figure*}
    \centering
    \includegraphics[width=\textwidth]{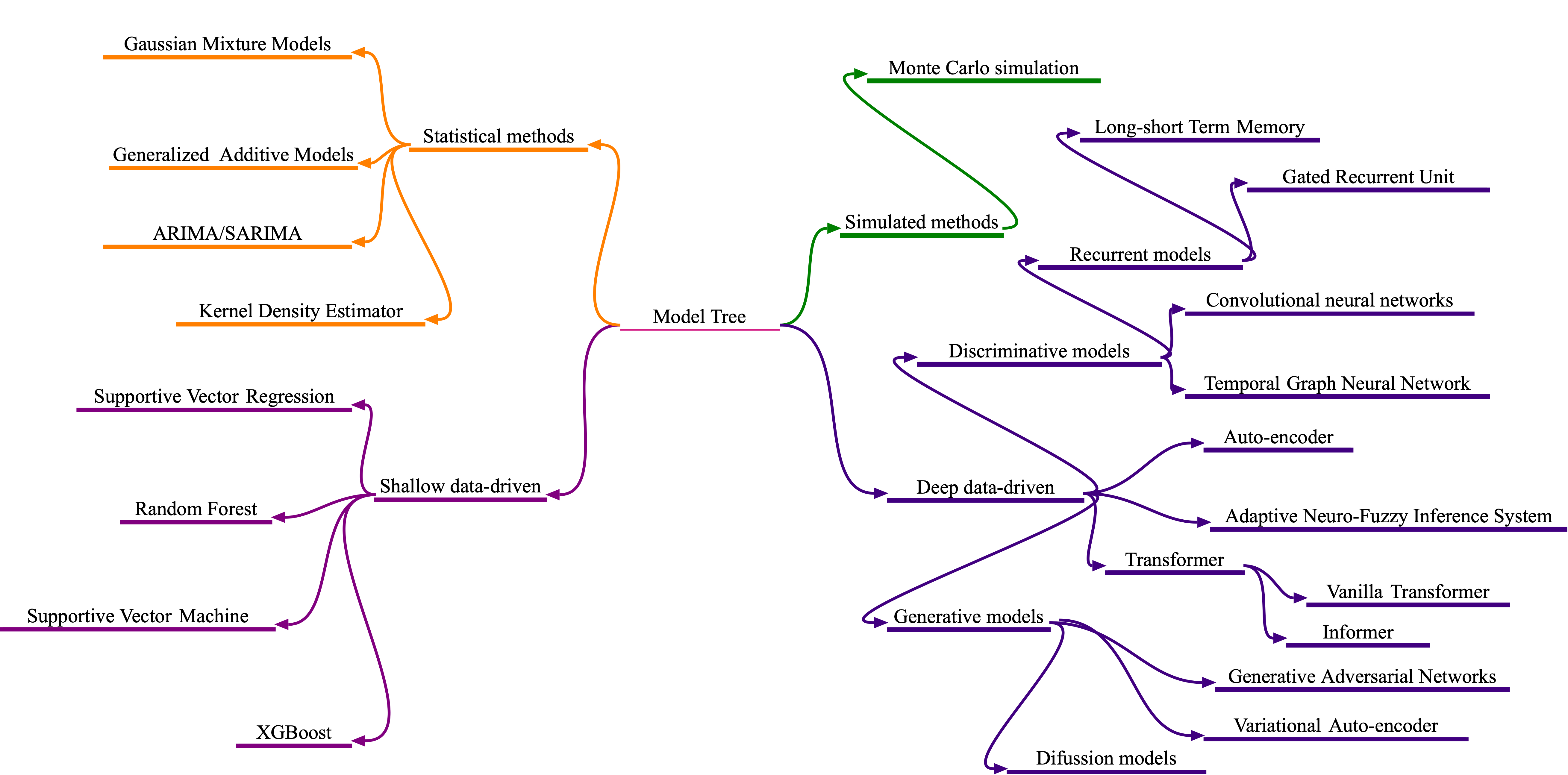}
    \caption{The model tree about adopted EV charging load modeling methods. \color{red}{TODO}}
    \label{fig:modeltree}
\end{figure*}
\subsection{Statistical Methods}
Statistical methods in EV charging modeling involve modeling the joint probability distribution of charging sessions. This approach allows for characterizing relationships between different variables that influence charging demand, i.e., arrival time, start charging time, and departure time. By modeling the joint probability distribution, statistical methods can capture dependencies among variables, leading to more accurate load forecasting. Common statistical methods employed in EV charging load modeling include regression analysis, stochastic modeling, and time-series analysis, which can be parametric or non-parametric. Parametric methods are those making assumptions about the underlying distribution of the data and estimating parameters of that distribution to make predictions, typically include Regression models (i.e., Linear Regression (LR), Auto-Regression (AR), Poisson Regression (PR)), Gaussian Mixture Models (GMM), and time-series models like Autoregressive Integrated Moving Average (ARIMA) and Seasonal ARIMA (SARIMA). These methods are characterized by their use of specific functional forms or distributions to model the correlations between variables and EV load. Lahariya et al. \cite{lahariya2020defining} estimated the energy consumption by proposing a GMM-based synthetic data generator to produce EV charging session samples, which implied temporal statistical modeling of EV arrivals and modeling departures in charging stations and their associated energy consumption. Autoregressive Integrated Moving Average (ARIMA) as a widely known statistical model, was designed to capture the temporal dependencies and patterns present in time-series and perform trend predictions based on past observations, was employed by \cite{kim2021forecasting, lu2021ultra, jin2020accurate} for EV charging load forecasting. Kim and Kim \cite{kim2021forecasting} conducted a comparative study that showed historical data demonstrating the potential for enhancing the ARIMA model's short-term and mid-term predictive capabilities. Jin et al. \cite{jin2020accurate} employed the GMM to analyze spatial information patterns, including the pile's electric orders, parking, and charging locations, and an ARIMA to forecast the temporal section. In \cite{lu2021ultra}, Lu et al. introduced a Gaussian filter as a linear smoothing filter for eliminating Gaussian noises to facilitate ARIMA for regional charging load forecasting, which validated that the Gaussian filter can improve the model performance with low data density. 

The Generalized Additive Model (GAM) is a representative of the semi-parametric statistical model, where a semi-parametric model refers to the method that uses parametric components to capture global trends or patterns in the data but allows the flexible non-linear relationship between the predictors and the response variable. In \cite{amara2022benchmark}, Amara et al. presented a comparative benchmark to forecast the charging load and occupancy of EV charging infrastructures modeling from both direct and bottom-up ways, which employed Poisson regression, auto-regression, mixture regression, and GAM from the statistical aspects. This study indicated that direct modeling is better than bottom-up, and the direct approaches can be improved by adaptive regression strategy. 

However, the probabilistic assumptions for parametric models can lead to limitations in application fields. For instance, the assumption of Gaussianality can limit GMM application in non-Gaussian datasets. To avoid this drawback, earlier studies, as the two earlier studies in \cite{chung2018electric, khaki2019probabilistic}, scholars introduced Kernel Density Estimators (KDE) as a non-parametric method to EV charging modeling to address the mentioned issue of the parametric approach. In recent years, Chen et al. utilized the Ternary Symmetric KDE to estimate charging behavior like arrival time, staying duration, and charging capacity of EVs in the working area \cite{chen2020modeling}. 

\subsection{Simulated Methods}

Monte Carlo Simulation (MCS) is a widely used technique for modeling EV charging processes by simulating the behavior of individual EVs or charging stations. This method enables the prediction of aggregated charging demand by capturing real-world variability and uncertainties through repeated sampling. By aggregating the effects of diverse behaviors and conditions, MCS offers valuable insights into cumulative charging loads and their impact on the electric grid.

\begin{figure*}
    \centering
    \includegraphics[width=\linewidth]{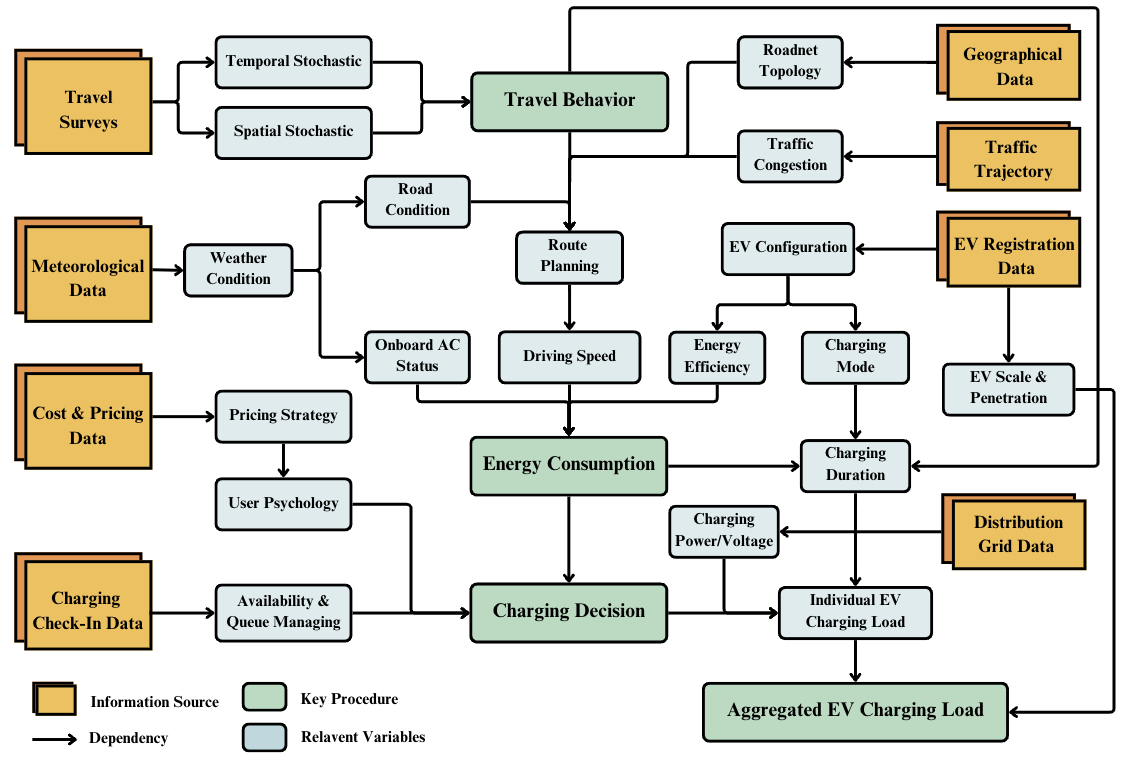}
    \caption{Logic Flow of Monte Carlo Simulation for Micro-Grid EV Charging Load. This flowchart illustrates the logical dependencies within the simulation process, with each arrow representing the flow of information or dependency between components. This flow is summarized based on literature~\cite{helmus2020data, roadnet_powergrid, roadnet_whether, han2020ordered, fuzzy4, spatio_temporal_roadnet, spatio_temporal_roadnet1, spatio_temporal_traffic, moghanlou2023probabilistic, bib1, bib3, long-shenzhen, zhang2021ev, zhang2020daily, roadnet_roadcond, liu2020electric, ni2020methodology, tayyab2021infrastructure, xing2022multi, bian2022multi, tian2022electric, gou2021charging, bib0, williams2024driving, zhuang2022load, zhang2023battery, chen2023spatio, liu2024electric}.}
    \label{fig:MonteCarlo}
\end{figure*}

MCS serves as a bottom-up simulation method, integrating data from various sensors and sources to create a comprehensive foundation for load management strategies. The key components incorporated in the simulation include vehicle travel patterns, user preferences, traffic conditions, weather influences, expert insights, and societal or environmental factors. Figure \ref{fig:MonteCarlo} illustrates the interdependencies involved in simulating aggregated EV charging loads. 

The MCS process can be divided into four distinct steps:
\begin{enumerate}
    \item \textit{Modeling travel patterns:} Identifies the movement behavior of EVs, including routes, distances, and frequency.
    \item \textit{Estimating energy consumption:} Calculates the energy needed for each trip based on the vehicle’s characteristics and driving conditions.
    \item \textit{Determining charging decisions:} Evaluates when, where, and how EVs charge based on user preferences, station availability, and tariffs.
    \item \textit{Calculating the aggregated charging load:} Combines individual charging events to predict the total load and its impact on the grid.
\end{enumerate}
These procedures link various parameters based on domain knowledge. Each parameter varies with time or conditions and can be modeled by
stochastic processes. This decision-level information fusion integrates human expertise, social knowledge, and physical laws into the simulation, creating a robust system for EV load modeling.

While Monte Carlo Simulation offers flexibility and comprehensive modeling capabilities, it also faces several limitations. A key drawback lies in the oversimplification of complex real-world variables, which can undermine the simulation's accuracy. For example, static assumptions, such as treating vehicle models uniformly, do not account for the diversity of EV models, market shares, or their evolution over time. Similarly, the rapid development of EV charging standards is often overlooked. For instance, CHAdeMO (a DC fast-charging standard) is being phased out in favor of the Combined Charging System (CCS), which offers higher power capacities. However, simulations frequently fail to incorporate these evolving standards due to the increased complexity they introduce.

Another challenge is that the more features integrated into the simulation, the more complex the interactions between them, leading to higher computational and design costs. Simulations that rely heavily on manually defined parameters and professional knowledge may result in underutilized data and reduced adaptability to real-world conditions. The trade-off between model complexity and computational feasibility remains a persistent issue in MCS-based EV load modeling, limiting its effectiveness for large-scale or rapidly evolving systems.

Monte Carlo Simulation provides a powerful framework for modeling EV charging behavior, offering flexibility to explore uncertainties and variability in load forecasting. However, the reliance on static assumptions and the complexity of integrating dynamic, evolving parameters can limit the accuracy and applicability of the simulation results. Despite these challenges, MCS remains a valuable tool in EV load modeling, especially when combined with expert knowledge and complementary modeling techniques.

The advantage of Monte Carlo simulation is that it can sufficiently analyze user behavior and be customized based on different focuses and modeling purposes. However, the drawback of this approach is that the sampling is performed based on many distribution assumptions, which, in some cases, is not realistic. Some literature aims to overcome the predefined stereotypical expectations, such as in \cite{helmus2020data}, Helmus et al. propose a dual-step clustering method to provide charging topology groups by user types. However, the limitation still exists. Moreover, with the increased amount of modeling features, the generating process is increasingly computationally expensive, which is ineffective in large-scale datasets. Furthermore, since the model setting for the Monte Carlo usually takes fixed values for parameters like battery capacity, power consumption per km, etc., complicated cases require multiple scenarios as model inputs, which can lead to robustness issues and existing gaps with the real-world situation. Meanwhile, the simulation results from MCS have difficulties in examination and evaluation due to the lack of clear objectives and formula determination.

\subsection{Data-driven Discriminative Methods}

\subsubsection{Machine Learning Models}

Machine Learning (ML) models for EV charging typically feature simple architectures, focusing on mapping input features to outputs without modeling the underlying probability distribution of the data. Common examples include Support Vector Machines (SVM)~\cite{vapnik1999nature}, K-means clustering, and decision trees. These models, often referred to as "shallow" models, have limited parameters compared to deep neural networks.

Shallow models are widely used in practical applications such as estimating the probability distribution function of EV charging stochastic~\cite{gao2021data} and forecasting charging demand~\cite{ge2020data, deb2022prediction}. Their ease of implementation and computational efficiency make them suitable for real-time applications. Furthermore, their simplicity ensures interpretability, helping stakeholders understand the influence of key factors, such as user behavior, time-of-use tariffs, and traffic conditions, on charging demand. Models like decision trees and K-means clustering can also capture non-linear relationships and identify patterns, providing insights into peak charging times and station utilization rates.

Despite their advantages, shallow models struggle with complex scenarios, limiting their use in advanced research~\cite{marzbani2023hybrid}. As a result, they are often employed in comparative studies or used as benchmarks to assess the performance of more sophisticated models. In some cases, they are enhanced with feature extraction techniques to improve their effectiveness~\cite{jahromi2022probability}.

Recent studies have integrated shallow ML models into more complex algorithms to further improve performance. For instance, Gang et al.~\cite{gang2021load} combined a random forest with a generative model for load forecasting at various charging stations. Wang et al.~\cite{featuremining_GCN_GRU} used K-means clustering to group spatial regions with similar Points of Interest, while Deng et al.~\cite{deng2023asa} employed a single-class SVM for abnormal detection tasks. These examples demonstrate how shallow models can effectively contribute to hybrid and ensemble approaches in EV charging applications.

\subsubsection{Temporal Dependency Models}
Temporal dependency models in EV charging capture and predict patterns in charging demand over time by learning sequential relationships within time-series data, enabling accurate forecasting and effective energy management. Long Short-Term Memory (LSTM)~\cite{hochreiter1997long} is one of the most widely used models in time-series forecasting challenges and is extensively applied in predicting EV charging load which involves complex temporal patterns. Chang et al.~\cite{chang2021aggregated} developed an LSTM-based framework to forecast aggregated power demand from multiple fast-charging stations, addressing challenges posed by fluctuating power demand due to short charging sessions and high-rated power chargers. Similarly, Unterluggauer et al.~\cite{unterluggauer2021short} proposed a multivariate LSTM model for forecasting EV charging loads over 1-hour and 24-hour horizons.

Incorporating probabilistic methods, Zhou et al.~\cite{zhou2022using} combined LSTM with Bayesian probability theory to capture uncertainty in forecasts. Lu et al.~\cite{lu2021ultra} further improved LSTM by integrating it with a Gaussian filter for linear smoothing, demonstrating that LSTM outperforms Gaussian-ARIMA models, especially when data density is low. Shanmuganathan et al.~\cite{shanmuganathan2022deep} introduced a framework combining Empirical Mode Decomposition (EMD) with the Arithmetic Optimization Algorithm (AOA) to forecast demand based on decomposed intrinsic mode functions. Feng et al.~\cite{feng2021load} developed a hybrid model integrating the first-order residual correction grey model (EMGM) with LSTM, where EMGM provides initial predictions and LSTM periodically corrects forecasting errors. Aduama et al.~\cite{featurefusion-lstm} enhanced LSTM with multi-feature fusion to improve forecast accuracy.

Researchers have also explored LSTM variants to address spatio-temporal forecasting challenges. For instance, Mohammad et al.~\cite{mohammad2023energy} proposed architectures using ConvLSTM and BiConvLSTM encoders with LSTM decoders to capture both spatial and temporal patterns from EV demand data. LSTM has also been utilized in ensemble models to boost forecasting performance. Yin et al.~\cite{lstm_xgboost} introduced an LSTM-XGBoost ensemble model for dynamic forecasting based on orderly EV charging behavior, while Huang et al.~\cite{huang2020ensemble} developed an ensemble framework combining predictions from artificial neural networks (ANN), recurrent neural networks (RNN), and LSTM.

The integration of Convolutional Neural Networks (CNN) with LSTM has shown promise in capturing both spatial and temporal dependencies. While CNNs effectively extract spatial features, their ability to capture temporal patterns is limited by the fixed receptive field of convolutional kernels. To address this, Zhou et al.~\cite{zhou2022ev} proposed a CNN-LSTM model for station-level charging load forecasting. Zhang et al.~\cite{zhang2022short} extended this approach with a multi-channel CNN (MCCNN) for multi-time-scale feature extraction, combined with a Temporal Convolutional Network (TCN) to capture temporal dependencies, outperforming models such as ANN, LSTM, CNN-LSTM, and TCN.

In addition to LSTM, the Gated Recurrent Unit (GRU)~\cite{cho2014learning}, a simplified variant of LSTM, has been applied to address time-series forecasting problems. GRU mitigates the vanishing gradient problem while requiring fewer parameters, making training more efficient. Guo et al.~\cite{guo2021short} optimized GRU parameters using a genetic algorithm (GA) and applied the optimized GRU for short-term EV load forecasting in North China with multisource data. GRU has also been integrated into generative models to enhance their ability to learn temporal relations, as demonstrated by Shen et al.~\cite{shen2022short}.

\subsubsection{Spatio-Temporal Graph Models}
\begin{figure*}[!htb]
    \centering
    \includegraphics[width=\textwidth]{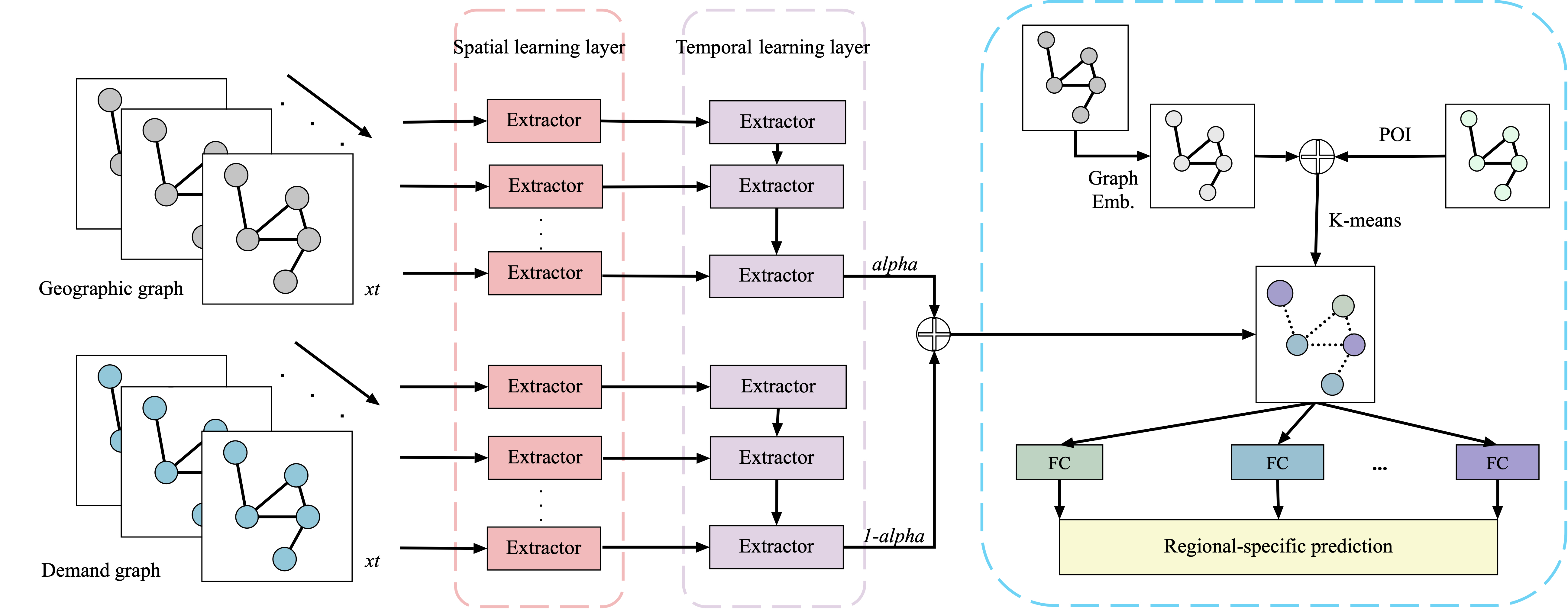}
    \caption{Sample architecture of graph-based spatio-temporal EV charging modeling. This architecture \cite{featuremining_GCN_GRU}, typically employs a sequential combination of spatial and temporal learning layers to extract spatio-temporal features from graph-structured input data.}
    \label{fig:TemporalGCN}
\end{figure*}

Graph Neural Networks (GNN) is another approach that provides flexibility in modeling spatial features in a grid. Given that the distribution grid is usually formed as complex topological and irregular structures, graph-based models are suitable for regional demand modeling compared to CNN, which is better at handling regular data. To enhance the model performance in modeling not only spatial distribution but also temporal dependencies, Temporal Graph Convolutional Network (T-GCN) \cite{zhao2019t}, as an extension of GCN that accounts for a temporal signal by combining external LSTM layers, was employed to model spatial and temporal features with raster maps for short-term forecasting purposes (daily, weekly, and monthly) by Huttel et al. \cite{huttel2021deep}. Li and Qi \cite{li2023ev} proposed an integrated Edge Aggregation Graph Attention Network (EGAT) with LSTM to capture the line renovation accompanied by adjacency relationship change and renew the node characteristic through attention scores. This study comprehensively considered the impacts of the infrastructure upgrade and modeled the line renovation and the changes in the topology structure. In the latest work by Wang et al. \cite{featuremining_GCN_GRU}, the GCN-GRU structure was employed to learn the geographical graph and demand graph on the grid, which well-described the location and corresponding demand of charging points within the distribution network. In \cite{luo2021deep}, the spatio-temporal GCN forecasts the charging station availability. The fundamental logic of these methods to enhance the model ability in spatio-temporal distribution learning followed the pattern of employing a graph-based spatial learning layer followed by a temporal learning layer ( i.e., GRU \cite{featuremining_GCN_GRU}, LSTM \cite{huttel2021deep}, Informer \cite{luo2023ast}). An example is visualized in Figure \ref{fig:TemporalGCN}.


\subsubsection{Transformer-based Models}
Besides, multi-head self-attention mechanism (Transformer) \cite{vaswani2017attention} was primarily introduced to capture the context dependencies in natural language processing and then was introduced to other fields such as computer vision \cite{wang2020non} due to its outstanding performance and efficiency in parallel data processing. The transformer was naturally introduced to the energy industry to capture the temporal dependencies in electricity load and model the future demand. In 2022, Hi et al. propose a Self-attention-based Machine Theory of Mind (SABMToM) to balance the EV charging load variation from historical data and current trends and achieve competitive forecasting results \cite{hu2022self}. Shi et al. \cite{shi2023spatial} realized that transformer-based models can accurately co-estimate the vehicle battery states and proposed a Bidirectional Encoder Representation from Transformers for Batteries (BERTtery) mechanism to estimate the state of charge (SoC). Another application of the Transformer was in \cite{li2023diffcharge}, where the authors embedded a multi-head self-attention layer in the denoising network to present the fused features sufficiently. However, the vanilla Transformer still faces challenges in long series time forecasting (LSTF) tasks. On the one hand, the quadratic time, memory complexity, and error accumulation caused by inherent architecture design lead to great limitations in handling large-scale tasks \cite{zhu2023time, zeng2023transformers}. On the other hand, the time-series data are in a longer scope than the NLP sentences, which have higher requirements for long-term dependencies capturing abilities. To address the research problem, Zhou et al. \cite{zhou2021informer} propose the Informer model with ProbSparse self-attention, replacing the autoregressive decoder manner as the generative decoder, achieving the competitive forecasting performance in LSTF tasks. Based on this work, a MetaProbformer that integrates the Informer with a meta-learning algorithm for optimal parameter initialization was proposed by Huang et al. \cite{huang2023metaprobformer}. The proposed probabilistic load forecasting framework is validated as owning better performance on both short-term and long-term tasks. Besides, Luo et al. propose an attributed-augmented spatio-temporal graph informer (AST-GIN) architecture to extract internal and external spatio-temporal dependencies in transportation data \cite{luo2023ast}. Research on the limitations of Transformers in LSTF problems is ongoing in the academic community. This process introduces many noteworthy new models as potential alternatives to Transformers. In the latest study \cite{liu2023itransformer}, Liu et al. propose an iTransformer that inverts the duties of the attention mechanism and the feed-forward network. As an alternative backbone in time-series forecasting tasks, the iTransformer achieves consistent state-of-the-art in real-world datasets and, at the same time, brings new opportunities to the application field of EV charging load forecasting. 

In addition, some alternative models are not the majority preferred by authors but existing inspired points in model employment. As in \cite{ahmed2023neuro}, Ahmed et al. utilized an Adaptive Neuro-Fuzzy Interface System (ANFIS) in conjunction with a Search and Rescue (SAR) framework to effectively simulate intricate dynamic energy emission dispatch scenarios. As a hybrid model, using ANFIS leveraged the good interpretability of traditional Fuzzy Systems and the flexible predictive capabilities of data-driven techniques. In \cite{luo2020load}, Luo et al. employed the Atacked Auto-Encoder Neural Networks (SAE) to perform short-term predictions on the charging station load at the edge platform based on the selected inputs, including historical load, weather type, temperature, day type, and charging station type. As an unsupervised learning model, SAE enabled the automatic discovery of the latent representations from the raw inputs without requiring manual feature engineering. \\

\subsubsection{Physics-Informed Neural Networks}
Physics-Informed Neural Networks (PINNs)~\cite{raissi2017physics} offer a novel approach for integrating scientific knowledge into neural networks, providing solutions to several limitations encountered in purely data-driven approaches. In the context of EV charging, relevant physics often involves the degradation dynamics of lithium-ion batteries and state-of-charge (SoC) estimation.

Tian et al.~\cite{tian2022battery} proposed a PINN-based deep learning framework for battery SoC estimation. Conventional neural networks typically focus on direct mappings between input variables and SoC outputs, often neglecting temporal dependencies. To address this, the authors enhanced a deep neural network by decoupling voltage and current sequences into open-circuit voltage, ohmic response, and polarization voltage for input augmentation. This approach was combined with a Kalman filter-based framework, integrating neural network predictions with Ampere-hour counting—a key physics-based principle—formulated as:
\begin{align*}
    z_k = z_{k-1} + \frac{\Delta t \eta}{Q} I_{k-1}
\end{align*}
where $\Delta t$ is the sampling interval, $\eta$ is the coulombic efficiency, $Q$ is the battery capacity, and $I$ is the current at time step $k-1$. This method ensures accurate SoC predictions by blending data-driven estimation with physical laws.

Beyond refining neural network predictions using physics, a popular approach in PINNs is to incorporate physics knowledge directly into the model’s loss function. PINNs modify the network loss by adding an extra ODE/PDE loss term, with the overall loss given by:
\begin{align*}
    \mathcal{L} = (1 - \alpha) \mathcal{L}_{NN} + \alpha \mathcal{L}_{\text{ode/pde}}
\end{align*}
where $\alpha$ controls the contribution of the physics-informed component. Wen et al.~\cite{wen2023physics} implemented this approach in their Deep Hidden Physics Model (DeepHPM) to explore battery degradation dynamics using a Verhulst dynamic model represented by a first-order ODE.

While much of the research focuses on battery SoC estimation, recent studies have applied PINNs to forecasting EV charging demand. Qu et al.~\cite{qu2023physics} developed a physics-informed, attention-based graph learning framework for regional EV charging demand forecasting. Their approach employed physics-informed meta-learning in the pre-training stage to capture adaptive charging behaviors driven by price fluctuations. Kuang et al.~\cite{kuang2024physics} extended this work by introducing a method to model the price elasticity of demand using PDEs, enhancing both demand forecasting and pricing strategies.

PINNs offer several advantages that make them highly suitable for EV charging load modeling. One of the key benefits is their ability to enhance accuracy by integrating physics-based knowledge directly into the model, which helps prevent overfitting and ensures physically consistent predictions. This is especially valuable in scenarios where conventional neural networks may struggle with capturing complex dynamics, such as battery state-of-charge (SoC) degradation or demand elasticity. Another advantage of PINNs is their ability to accelerate convergence during training. The inclusion of physical priors guides the learning process toward feasible solutions, reducing the need for extensive training iterations and improving computational efficiency. Moreover, PINNs generalize well, even in cases with limited data, since the embedded physical laws compensate for the lack of large datasets. This makes them particularly useful for modeling systems like battery degradation, where high-quality labeled data may be scarce. Additionally, PINNs naturally accommodate systems governed by differential equations, such as those found in battery dynamics and demand forecasting, providing robust modeling capabilities.

However, despite these advantages, PINNs also come with certain limitations. Developing and implementing PINNs requires significant domain expertise, as both the neural network design and the selection of appropriate physical models, such as ODEs or PDEs, must be carefully considered. This adds complexity to the model development process and can increase design effort and costs. Furthermore, the inclusion of physics-based constraints introduces additional computational overhead, leading to longer training times and increased resource requirements. Another challenge with PINNs lies in the sensitivity of their performance to hyperparameter tuning, particularly the balance between the data-driven loss and the physics-informed loss, controlled by the hyperparameter $\alpha$. Improper tuning of this parameter can result in under- or over-regularization, affecting the model's performance. Finally, the reliance on domain-specific knowledge may limit the broader adoption of PINNs, as identifying suitable physical laws for different applications requires expertise that may not always be available in interdisciplinary teams.
\subsection{Data-driven Generative Models}
Generative AI, a concept gaining recent popularity, functions primarily as a training strategy, empowering models to generate samples with a distribution similar to the original inputs. Due to their outperforming capability in capturing the inherent nature of data, generative models are typically employed to handle tasks such as data augmentation, including profile imputation and generation, as well as charging demand estimation in EV charging load modeling. 

These generative models are categorized based on their operational principles for capturing latent spaces. The first category involves mathematical approximation and includes Variational Auto-encoder (VAE) \cite{kingma2019introduction} and Denoising Diffusion Probabilistic Models (DDPM) \cite{ho2020denoising}. The second category, rooted in game theory, is exemplified by the Generative Adversarial Networks(GAN) \cite{goodfellow2014generative}. 

The VAE and DDPM have similar fundamental rules, aiming to model the approximated distribution from heterogeneous data explicitly. The difference is that the learned latent representation reconstructs the output of the VAE. In contrast, DDPM is derived through a reverse denoising process based on the gradually forwarding Gaussian-noise-adding procedure. In an earlier study, Pan et al. employed the VAE for charging load profile generation \cite{pan2019data}. In \cite{li2023diffcharge}, Li et al. explored the application of DDPM as the generator to achieve data augmentation for forecasting enhancement. As the two generative models based on mathematical approximation, the VAE and DDPM can provide reliable estimations with better interpretability than GAN. However, in real-world datasets, their performance is usually not as good as GAN.

Generative Adversarial Networks (GAN) employ a generator to translate the noise into fake samples that are as similar as possible to the real data and a discriminator to distinguish the fake samples from real data. The generator and discriminator are trained to play against each other until they reach a Nash equilibrium. Benefiting from the adversarial training strategy, GAN is showing remarkable performance in various application domains. With the widespread adoption of GAN in addressing EV charging modeling challenges, we conducted a comprehensive analysis and visualization of their application in imputation, generation, and prediction tasks within EV charging modeling.

\begin{figure*}[!htb]
    \centering
    \includegraphics[width=0.8\textwidth]{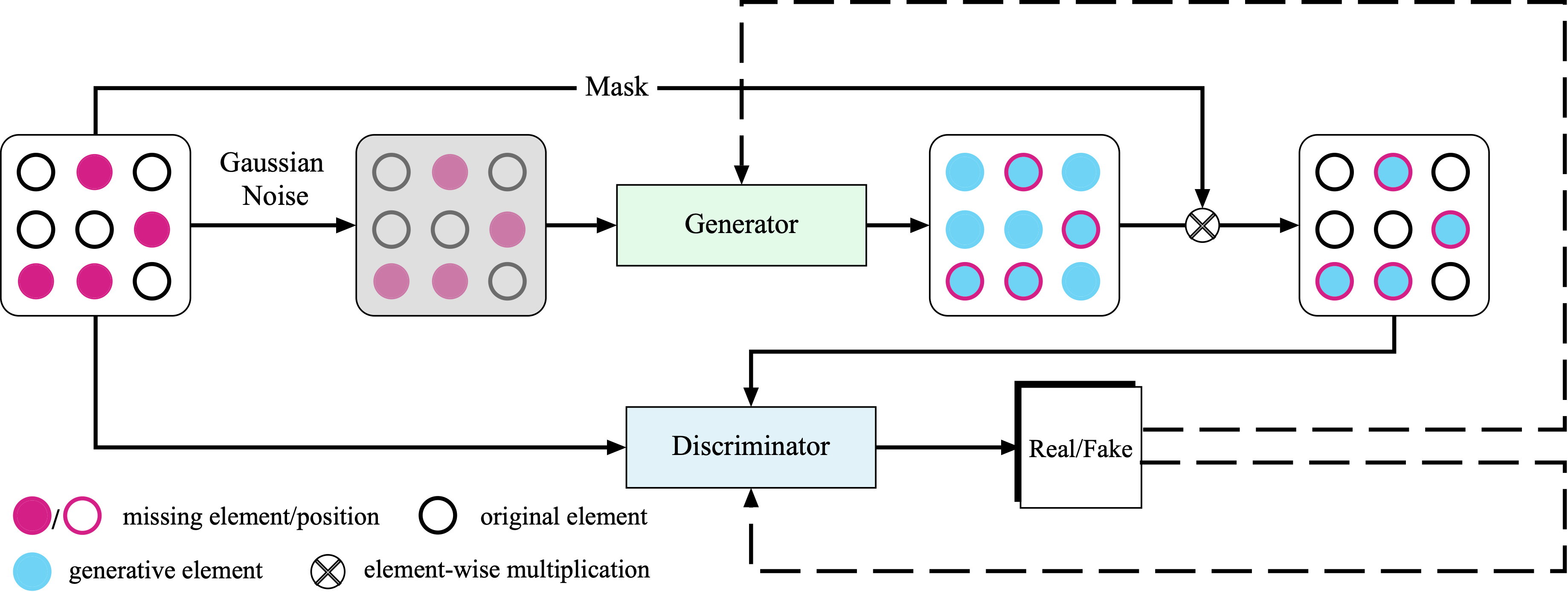}
    \caption{A general architecture for GAN in profile imputation. Extracted from Shen et al.'s work \cite{shen2022short}. The solid lines with the arrows represent the data flow directions, while the dashed lines with the arrows represent the backpropagation process.}
    \label{fig:gan_imputation}
\end{figure*}

The employment of GAN to address the imputation problem was illustrated in Shen et al.'s \cite{shen2022short} work. This work proposed a GAN-based imputation method for the time-series charging load data in low-quality situations. The GRU units are integrated into the generator and discriminator so that the GAN can better capture temporal features. A binary mask to indicate the coordinates of missing elements was applied to filter the generative data efficiently. Subsequently, a Mogrifier LSTM is employed to obtain the forecast results. The extracted training framework of this work is visualized in Figure \ref{fig:gan_imputation}. Besides, a coordinated dispatch strategy of EVs and thermostatically controlled loads (TCLs) model based on an improved generative adversarial network (IGAN) is proposed by Tao et al. in \cite{tao2021data}. The application of GAN in reconstructing EV charging load profiles effectively contributes to the training process less affected by the missing data.

GAN as a synthetic profile generator was found in \cite{gang2021load, huang2022multinodes, madahi2022overarching, forootani2023transfer}. In work by Gang et al. \cite{gang2021load}, a generative countermeasure network was proposed to generate spatio-temporal load matrices within specific regions. These generative outputs were then subjected to the min-max standardization process and utilized for predictions with a Random Forest model, leading to the acquisition of regional load profiles with reduced generative errors. In addition to generating time-series load profiles, GAN generates uncertainty profiles. As \cite{madahi2022overarching}, Madahi et al. take an unconventional approach by eschewing the typical methods of sampling model characteristics from predefined distributions or establishing probabilistic models. Instead, they use GAN to generate data for variables such as ambient temperature, solar irradiation, system power, daily mileage, and arrival and departure times. This comprehensive approach considers multiple dimensions and model parameters for defining the architecture. Besides, to differentiate the distribution under various characteristic conditions, Zhang et al. \cite{zhang2022multi} introduced a Conditional Wasserstein GAN (CWGAN) for EV charging parameters data generation with conditional labels to distinguish charging load in office area, business area, and residential area. The training goal was set to reach the stability of Wasserstein distance, and the KDE estimated results validated the alignment between the original and generated data.

\begin{figure*}[!htb]
    \centering
    \includegraphics[width=0.8\textwidth]{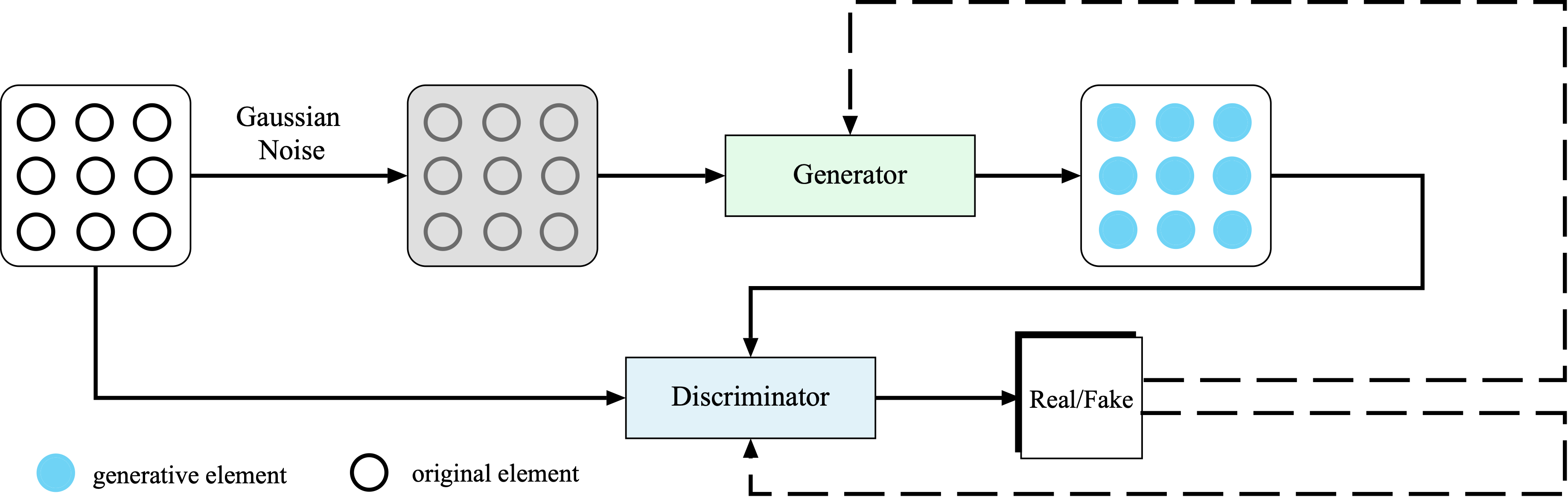}
    \caption{A general architecture for GAN in profile generation. Extracted from studies \cite{huang2022multinodes, zhang2022multi, gang2021load, madahi2022overarching}. The solid lines with the arrows represent the data flow directions, while the dashed lines with the arrows represent the backpropagation process.}
    \label{fig:gan_generation}
\end{figure*}

\begin{figure*}[!htb]
    \centering
    \includegraphics[width=0.8\textwidth]{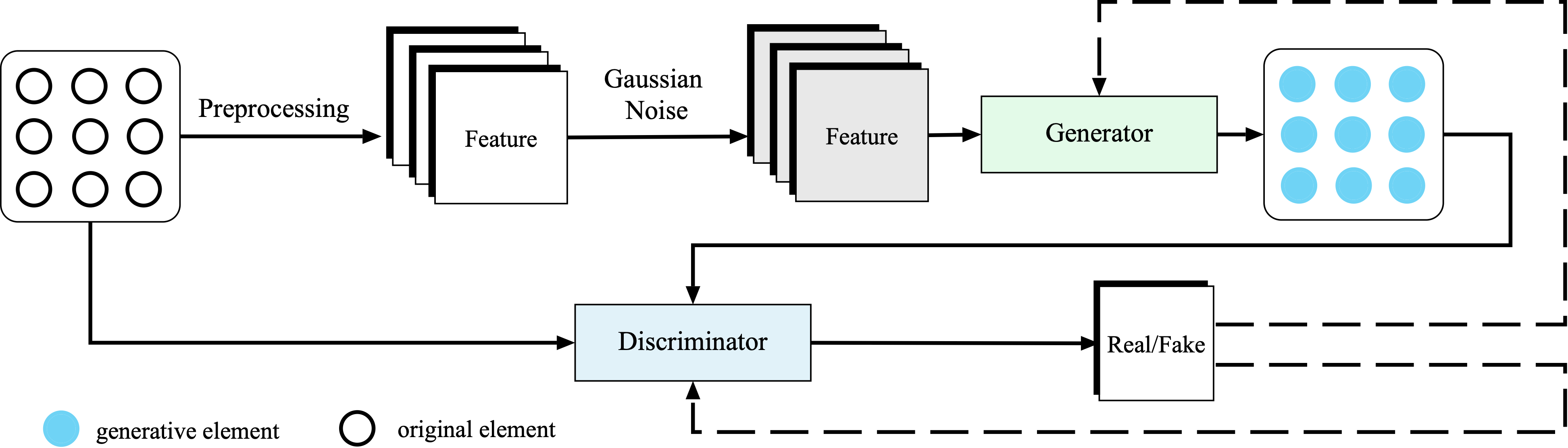}
    \caption{A prediction architecture for GAN in profile generation. Extracted from studies \cite{huang2022multinodes, gu2023time}. The solid lines with the arrows represent the data flow directions, while the dashed lines with the arrows represent the backpropagation process.}
    \label{fig:gan_prediction}
\end{figure*}

Moreover, scholars explored GAN's application in forecasting future EV charging trends. 
Figure \ref{fig:gan_prediction} visualized a general architecture design for GAN in prediction tasks extracted from \cite{huang2022multinodes, gu2023time}. 
Huang et al. \cite{huang2022multinodes} utilized the 2D correlations to construct the training profiles based on correlation coefficients and sequentially employed a Wasserstein GAN with gradient penalty (WGAN-GP) to generate the interval forecasting on the target day, which is named multi-node multiple-correlation-day joint charging scenarios (GMDJS). In a later work, Gu et al. proposed a time-series Wasserstein GAN (TS-WGAN), which contained a transformer-based generator for sequential global relations learning in a longer time interval and a CNN-based discriminator for local information capturing to estimate the EV battery SoC. In \cite{gu2023time}, Gu et al. propose a time-series Wasserstein GAN (TS-WGAN-GP) to forecast the univariate battery SoC. This model employs a transformer-based generator for sequential global relations learning over time and a CNN-based discriminator for local information capturing.

\subsection{Large Language Models}

Large Language Models (LLMs) have emerged as transformative tools in data mining, demonstrating exceptional abilities in natural language processing (NLP), predictive analytics, and pattern recognition. Models like \texttt{GPT-4} are trained on extensive datasets, enabling them to generate coherent, context-aware text and uncover complex patterns within unstructured data. Although LLMs have proven valuable in many fields, their application in the electric vehicle (EV) domain remains relatively nascent. However, LLMs hold substantial potential to revolutionize EV charging load modeling through their ability to synthesize, analyze, and predict using diverse data sources.

In the context of EV charging, LLMs can integrate historical charging data, real-time grid information, weather forecasts, and user behavior to deliver more accurate predictions of charging demand. These models can analyze patterns across multiple time horizons and adjust predictions dynamically, supporting proactive energy management. For example, Chaturvedi et al.~\cite{chaturvedi2023generative} leveraged GPT-4 to solve challenges related to price discovery in distributed digital systems, showcasing the adaptability of LLMs for predictive tasks in complex environments.

Moreover, LLMs can enhance decision-making by extracting insights from unstructured data sources such as maintenance logs, user feedback, and regulatory documents—resources that are often underutilized in traditional models. This ability to process textual and qualitative data opens new opportunities for improving charging infrastructure planning and user experience optimization.

An example of LLM integration in this field is demonstrated by Feng et al.~\cite{feng2024large}, who introduced an LLM-based agent framework for simulating EV charging behavior. Their model incorporates modules such as Persona, Planning, Perception, Memory, Decision-Making, Reflection, Action, and Environment, each reflecting user preferences, psychological traits, and environmental factors. This framework illustrates how LLMs can personalize charging strategies, optimizing processes by considering both individual user needs and external factors.

By harnessing LLMs' ability to extract actionable insights from diverse datasets, EV charging load models can become more robust and adaptive. This integration promises to enhance energy management efficiency, optimize infrastructure deployment, and create personalized, seamless user experiences, ultimately contributing to the sustainable growth of the EV market.

\section{Challenges and Opportunities}
\label{sec:discussion}
\begin{table*}[!htb]
    \centering
    \footnotesize
    \vspace{0.2cm}
    \begin{tabularx}{\textwidth}{l|X|X|X}
    \toprule
         & \textbf{Physical Principles} & \textbf{Practical Experience} & \textbf{Digital Data} \\
         \midrule
         \textbf{Statistical} & Typically not explicitly incorporate physical laws, but indirectly adhere to them. (i.e., the distributional assumptions of charging processes may align with physical principles without explicitly stated.) & Informs the selection of variables and features in statistical models. (i.e., selecting which parameters are the input to jointly determine the PDF of charging load.) & Heavily rely on historical charging data to identify patterns, trends, and relationships between variables. Data collected from the dispatching network participants and external sources are used to train and evaluate the model. \\
         \midrule
         \textbf{Simulated} & Simulated models directly incorporate physical laws governing the behavior of electrical circuits, battery chemistry, heat dissipation, and other relevant factors. These laws accurately simulate the charging process, ensuring the model reflects real-world physics. & Common sense informs the design and validation of simulated models. For example, common-sense knowledge about the limitations of charging infrastructure, the capacity of electrical grids, and user behavior guides the development of realistic simulation scenarios. & Simulated models may utilize numerical data to perform statistical characterization for input stochastics. Real-world data can help validate simulated outcomes and ensure the accuracy of the simulation. \\
         \midrule
         \textbf{Data-driven} & May indirectly incorporate physical laws by selecting features and the training process. (i.e., features derived from physical knowledge may be included in the model to capture relevant relationships.) Able to provide physics-constrained information to participate in guiding model training. & Guide the selection of input features, preprocessing steps, and model interpretation in data-driven approaches. Domain knowledge about EV charging systems, user behavior, and environmental factors informs the modeling process and helps ensure the model aligns with real-world expectations.& Rely heavily on support from large-scale, high-quality datasets to train/validate/evaluate the machine/deep learning models. Feature engineering techniques may be used to extract relevant information from the data and improve model performance. \\ 
         \bottomrule
    \end{tabularx}
    \vspace{0.2cm}
    \caption{A summarization of the interactions between three information sources (physical knowledge, engineering experience, and open data) and three EV charging modeling approaches (statistical, simulated, and data-driven methods). }
    \label{tab: methodandinfo}
\end{table*}

This survey investigates three primary types of EV modeling methods: statistical, simulated, and data-driven approaches and their integration with physical principles, practical experience, and digital data. Table \ref{tab: methodandinfo} briefly analyzes the three types of information that support EV charging differentiated by modeling tools. Statistical and simulated methods are deeply intertwined with physical principles and domain knowledge, providing robust frameworks grounded in strong statistical assumptions. However, these methods often struggle to accurately capture the complexity of human behavior, which varies significantly under different economic, natural, and contextual conditions. In contrast, data-driven methods offer flexibility and adaptability but lack the inherent constraints provided by real-world physics, which can lead to the risk of overfitting. Additionally, the efficacy of data-driven models is heavily contingent on the quantity and quality of the available data, making them sensitive to data-related limitations.

Moreover, current studies predominantly rely on well-processed datasets, bypassing the challenges associated with handling real-world data irregularities and quality issues. However, addressing these challenges is essential for developing robust models that operate effectively in real-world conditions where data may be incomplete, noisy, or heterogeneous. Additionally, it is necessary to integrate more resources and multimodal information into the systems to enrich the data supporting EV charging behavior modeling. This integration, however, introduces challenges related to system alignment and modeling across diverse data sources. Future research should, therefore, prioritize methodologies for preprocessing and integrating diverse data sources, aiming to enhance the resilience and applicability of EV charging models in varied and unpredictable environments.

In EV charging modeling, the distinction between black-box and white-box methods is pronounced. White-box methods are grounded in domain insights and physical principles, whereas black-box methods derive their insights from data. Each approach has its own strengths and limitations: white-box methods provide transparency and are driven by established knowledge but may struggle with the complexity of real-world data; black-box methods excel in adaptability and pattern recognition yet often lack the interpretability and constraints offered by physical laws. Current studies on EV charging modeling often fall short of employing comprehensive and integrative methods that bridge these two paradigms. There is a notable gap in the development of gray-box models, which synergize domain knowledge with data-driven techniques, leveraging the strengths of both to create more robust and effective EV charging models. Future research should focus on advancing gray-box methodologies to overcome the individual limitations of black-box and white-box approaches, enhancing the overall accuracy and applicability of EV charging models.

In addition, EV charging modeling faces significant challenges in terms of model transferability and generalization. Existing studies primarily focus on load management or regulation issues within specific areas or regions. This localized approach often limits the applicability of these models to other contexts. However, for real-world applications, there is a critical need for models that can be directly deployed in various locations without the necessity for retraining. Achieving such versatility requires developing models that are accurate, robust, and adaptable to diverse conditions and datasets. This highlights the need for research efforts aimed at enhancing the transferability and generalization capabilities of EV charging models, ensuring their effectiveness across different environments and scenarios.

From the IoT perspective, EV charging modeling faces significant challenges related to privacy preservation and the effective utilization of IoT network information. As EV charging systems increasingly rely on IoT devices to collect and transmit data, ensuring the privacy and security of this data becomes paramount. Sensitive information, such as user identities, charging habits, and location data, must be protected from unauthorized access and potential breaches. Concurrently, the vast amount of data generated by IoT networks presents an opportunity for more accurate and efficient EV charging models. However, leveraging this information to its full potential requires sophisticated data integration and processing techniques that can handle the heterogeneous and dynamic nature of IoT data. Balancing the need for privacy with the need to utilize IoT network information effectively is a critical challenge that must be addressed to advance EV charging modeling.

In general, EV charging modeling encompasses a broad spectrum of downstream demands, including forecasting, classification, anomaly detection, and optimization. Each of these tasks presents unique challenges and requires specific approaches tailored to different objective goals. For instance, the optimization targets can vary significantly based on whether the focus is on minimizing charging time, reducing costs, enhancing grid stability, or maximizing the use of renewable energy. Existing studies tend to concentrate on a narrow set of predefined tasks, designing models specifically for those purposes. This approach often leads to siloed solutions that lack versatility and broader applicability. Consequently, there is a significant opportunity to develop a universal model capable of addressing a wide range of tasks within the EV charging domain. Such a model would integrate various objectives, providing a comprehensive solution that adapts to different scenarios and requirements. Advancing toward this goal would involve creating flexible, multi-functional models that can seamlessly transition between tasks such as forecasting future charging demands, classifying usage patterns, detecting anomalies, and optimizing charging processes, thereby significantly enhancing the efficiency and effectiveness of EV charging systems.

\section{Conclusion}
\label{sec:conclusion}
In this paper, we conducted a comprehensive literature review of information fusion in EV charging load modeling. This survey crosses the time scope from 2020 to 2024 and involves statistical, simulated, machine learning, and deep learning methodologies. The information fusion involved in modeling processes was discussed from data, model, and strategy insights. By synthesizing findings from data aggregation, model development, and training framework design, we illuminate key areas for future research and innovation in this dynamic field.




%

\onecolumn
\appendices

\twocolumn

\ifCLASSOPTIONcaptionsoff
  \newpage
\fi



\bibliographystyle{IEEEtran}
%

\bibliography{biblo}




%








\end{document}